\documentclass[jkps,preprint,fleqn,showkeys]{revtex4}
\usepackage[pdftex]{graphicx}
\usepackage{amssymb}
\usepackage{amsmath}
\usepackage{bm}
\usepackage{xcolor}

\begin{document}
\setcounter{page}{0}

\title[]{NaI(Tl) crystal scintillator encapsulated in two organic-scintillator layers with pulse shape data analysis}
\author{Jinyoung \surname{Kim}}
\author{Yujin \surname{Lee}}
\email{harubyeol126@cau.ac.kr}
\author{Byoung-cheol \surname{Koh}}
\author{Chang Hyon \surname{Ha}}
\email{chha@cau.ac.kr}
\affiliation{Department of Physics, Chung-Ang University, Seoul 06974, Republic of Korea}
\author{Byung Ju \surname{Park}}
\affiliation{IBS School, University of Science and Technology (UST), Daejeon 34113, Republic of Korea}
\affiliation{Center for Underground Physics, Institute for Basic Science (IBS), Daejeon 34126, Republic of Korea}
\author{In Soo \surname{Lee}}
\affiliation{Center for Underground Physics, Institute for Basic Science (IBS), Daejeon 34126, Republic of Korea}
\author{Hyun Su \surname{Lee}}
\affiliation{Center for Underground Physics, Institute for Basic Science (IBS), Daejeon 34126, Republic of Korea}

\date[]{Received XX April 2022}

\begin{abstract}
  Thallium-doped sodium iodide (NaI(Tl)) crystals are widely used in radiation detection applications, from gamma-ray spectroscopy to particle dark matter searches. However, if the crystal is exposed to relative humidity of even a few percent, its light emission degrades, making the crystal impractical as a detector. Surrounding the crystal with organic scintillators not only protects the surface of the crystal from humid air but also offers a new capability to tag backgrounds such as external gamma rays and surface contaminations. We developed a detector that is constructed by fully encasing a NaI(Tl) crystal in a plastic scintillator and then immersing the plastic--crystal assembly in liquid scintillator. Using data collected from this triple phoswich detector, a pulse shape analysis is able to identify the various radiation signals from the three scintillators. Additionally, we find that the crystal's emission quality is maintained for a month.
  
\end{abstract}


\maketitle

\section{Introduction}
Our universe is made of a large amount of non-radiative matter but its nature is not well understood.
Astrophysical measurements find that approximately 26\% of the energy in the universe is in the form of dark matter~\cite{Ade:2015xua}. In physics beyond the Standard Model, there is strong theoretical motivation for the nature of dark matter being a particle~\cite{1937ApJ86217Z,PhysRevD.98.030001}, a prime candidate that is known as the weakly interacting massive particle~(WIMP)~\cite{lee77}. Experimental searches for a WIMP have been performed in laboratories worldwide using a variety of methods~\cite{Schumann:2019eaa,Gaskins:2016cha,Kahlhoefer:2017dnp}. One traditional method among those searches looks for scintillation light produced by a nuclear recoil when a WIMP interacts in a target crystal~\cite{COSINE-100:2021xqn,Amare:2021yyu}. To date, there has been no direct measurement of a WIMP--nucleus interaction although there is a well-known anomaly reported by the DAMA (short for DArk MAtter) group, which records an annual variation in residual events from the background rate that could be attributable to the Earth's motion relative to the WIMP-filled galactic halo~\cite{Bernabei:2018yyw,Freese:2012xd}.

To test the DAMA's experiments use of thallium-doped sodium iodide~(NaI(Tl)) crystal is important.
A model-independent comparison using the same target material as DAMA's is crucial since experiments with other target materials have already ruled out the annual modulation signal at the same WIMP-nucleus crosssection region~\cite{XMASS:2018koa,XENON100:2015hgt}.
The NaI(Tl) target crystal is particularly interesting because it generally produces a large amount of light for a given amount of energy deposited and has relatively high stopping power in radiation detection~\cite{Adhikari:2017esn}. Because of the hygroscopic nature of the substance, however, it is challenging to formulate a suitable detector, and further development is required to achieve tight encapsulation against humidity~\cite{Choi:2020qcj}. Thus, an important indicator of a well-constructed detector is the stability of a high level of output light~\cite{LEE2020163141}.

In rare decay experiments, $^{210}$Pb contamination of the target crystal surface is typically observed, which limits the size of the fiducial region~\cite{AKERIB20171,XENON:2017fdb,HENNING201629,LEONARD2017169}. The $^{210}$Pb isotope has a relatively long half-life ($\rm T_{1/2}$ = 22 years), and the associated beta particles can mimic the dark matter interaction signal in the region of interest, particularly at low energies (below a few keV). To address both the effect of humidity and the radioactive contamination of the crystal surface, a phoswich encapsulation of a NaI(Tl) crystal can be used, with layers of organic scintillator serving as active vetoes. 

Studies show that signals from a crystal coupled with either a plastic scintillator~(PS) or liquid scintillator~(LS) can be distinguished from the signals from the outer layer organic scintillators by using a pulse shape discrimination method~\cite{sensors}. However, light quality of the crystal in this kind of the detector setup has degraded relatively quickly owing to the diffusion of humidity and to the incompatibility of the plastic and the crystal during molding of the soft plastic resin~\cite{LEE2020163141}. To address this problem, we developed a triple phoswich detector in which a crystal is encased first with PS and then with LS. This double-layer encapsulation further protects the crystal from humidity while allowing crystal signals to reach the photomultiplier tube~(PMT) photocathode, as the indices of refraction for the plastic and liquid are similar. Moreover, the three scintillation signals from the different materials are recorded simultaneously in a single PMT readout, making the active veto more effective.

Using the pulse shape discrimination method, the signal origins (from each individual substance or combination of substances) can be identified.
It is possible to distinguish them because the decay time of the NaI(Tl) crystal is 250~ns, whereas that of PS and of LS is 5~ns~\cite{knoll}.
In addition, as the three scintillators have different energy scales for the same deposited energy,
the combined signals (e.g., LS+crystal or PS+crystal) can be distinguished as well.
Thus, the detector is able to tag radiation events as external or internal according to whether
they produce light only in the organic scintillators or in coincidence with signals from the crystal.
This detector concept offers efficient background rejection and crystal protection and can be used in a wide range of applications, from environmental radiation monitors to medical devices. 

\section{Experimental setup}

A NaI(Tl) crystal is hygroscopic and brittle\footnote{It has a hardness of 2 on the Mohs scale (https://www.crystals.saint-gobain.com).} and therefore requires careful handling.
When attacked by humidity, the surface of the crystal is degraded in terms of its light emission quality.
It typically becomes more opaque, and crystal light would not easily emerge from the scintillator.

A small bare NaI(Tl) crystal\footnote{This crystal was grown at the Center for Underground Physics, Institute for Basic Science, Daejeon, Republic of Korea.}
was machined to a cylindrical shape of diameter 15~mm and length 18~mm.
The light yield of this particular crystal was previously measured as $17.1\pm0.5$~photoelectrons/keV
in a separate experiment. As a comparison, typical NaI(Tl) dark matter experiments report 5--15 photoelectrons/keV~\cite{Adhikari:2017esn}.
The crystal was then placed in a low-humidity glove box\footnote{The absolute humidity in the glove box was maintained below 100~ppm H$_2$O.} and hand-polished using lapping films until it displayed a clear surface.

\begin{figure}[!htb]
  \begin{center}
      \includegraphics[width=0.93\textwidth]{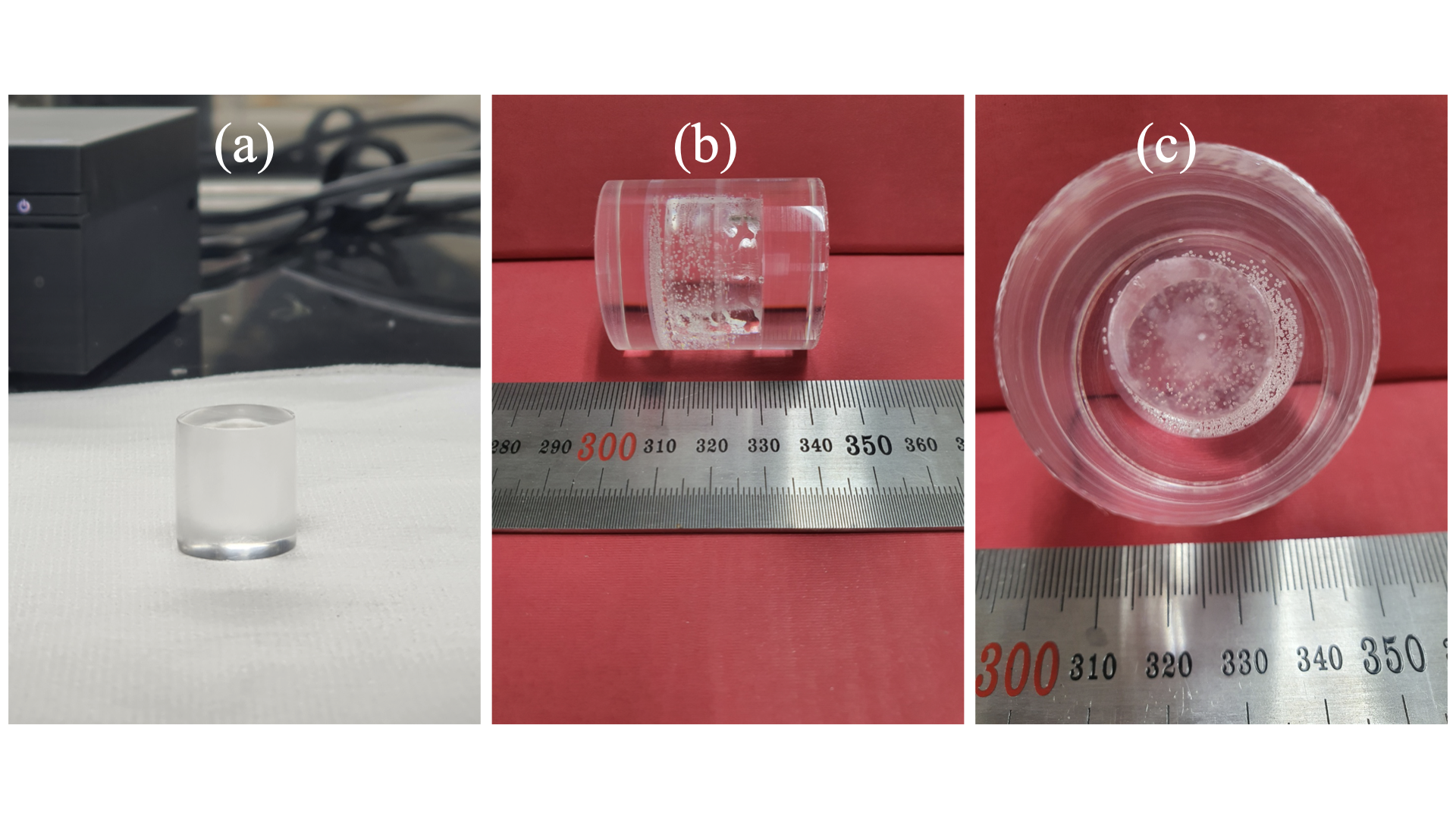}
      \includegraphics[width=0.93\textwidth]{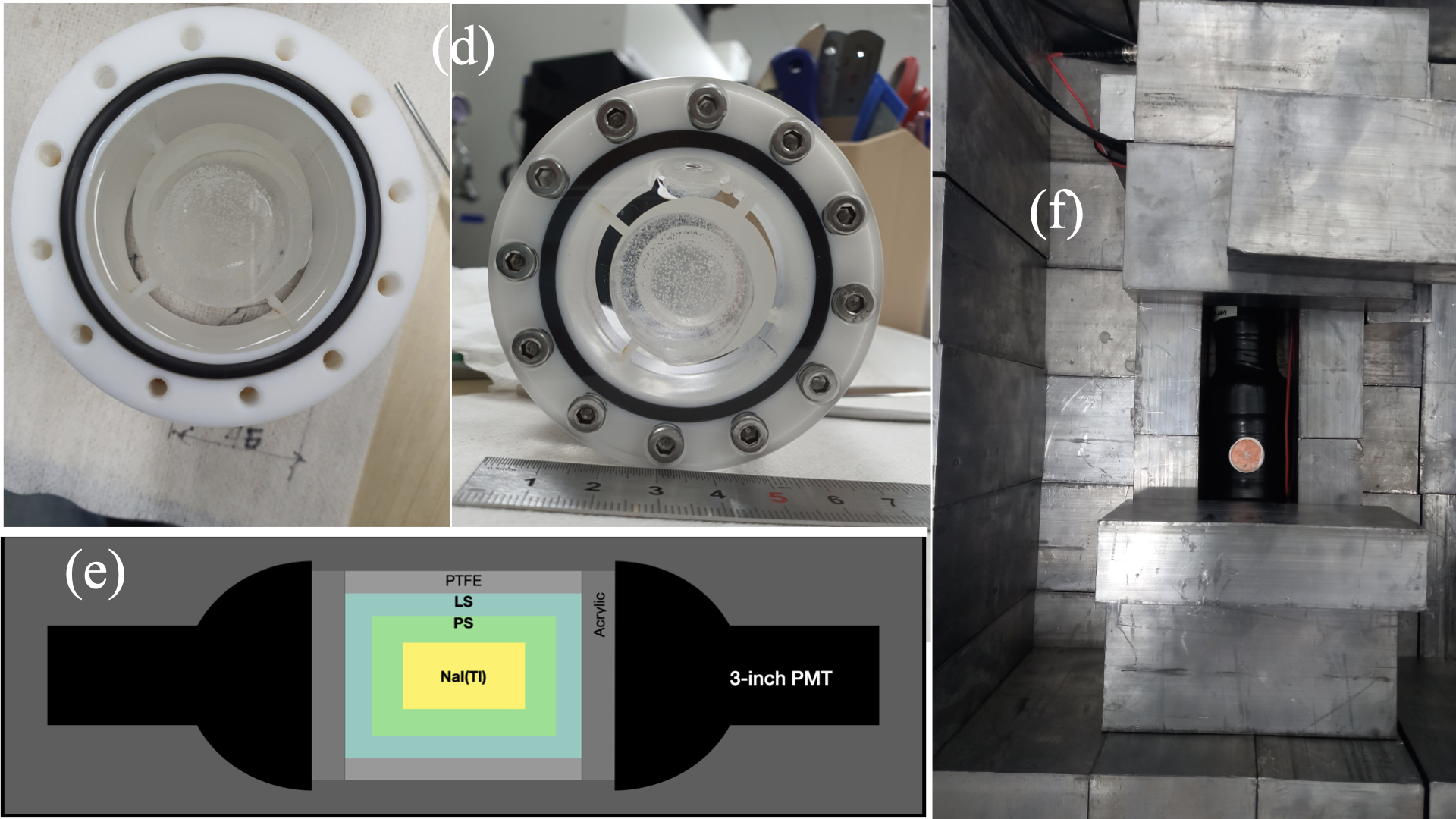}
  \end{center}
  \caption{NaI(Tl) crystal and plastic--liquid scintillator assembly. (a) Photo of bare crystal,
    (b),(c) photos of the plastic-encapsulated crystal assembly,
    (d) photo of the triple phoswich detector, consisting of the plastic scintillator (PS)--crystal assembly
    encompassed by liquid scintillator (LS)
    held in a polytetrafluoroethylene (PTFE) container, (e) schematic diagram showing the arrangement of the complete detector
    (with small support parts omitted for better visibility), and
    (f) photo of the final detector configuration placed in a lead-shielded dark box.
  }
  \label{naips}
\end{figure}

Next, we prepared a cylindrical container (diameter 17~mm and depth 20~mm) and
a lid, each made entirely of PS.
The crystal was placed in this container and
optically coupled by filling the gap between the crystal and PS
with optical cement\footnote{We used Saint-Gobain BC-600.}.
The lid was also attached using the optical cement.
After hardening, this PS--NaI assembly was machined and its surface polished further
until it had a diameter of 35~mm and length 42~mm, as shown in Fig.~\ref{naips}a, b.
Finally, the surface of the assembly was coated with the optical cement to prevent chemical reaction against other materials.
The assembly was inspected visually over the course of several days, and no degradation was observed.
However, the optical cement contracted in the process of hardening, creating some bubbles in one side,
which will increase the scattering of scintillation photons.

Separately, we prepared an open-ended cylindrical tube of 10-mm-thick polytetrafluoroethylene~(PTFE), with an inner diameter of 45~mm
and a length of 50~mm. On each end was an acrylic window, 
attached to the PTFE pipe using bolts and an O-ring.
The inner part of this second container also serves as a diffusive light reflector.
The PS--NaI assembly was positioned at the center of the container,
using small plastic support parts to stabilize it.
This container was filled with liquid scintillator~(LS), 
and two acrylic windows were used to light-couple and seal the final unit, forming the detector.
The photos of the crystal and detector assembly can be found in (a)--(d) of Fig.~\ref{naips}.
The completed experimental setup is shown in (e) and (f) of Fig.~\ref{naips}.

Properties of the LS counter have been widely studied in many experiments
requiring a detector having a large volume and/or an irregular shape~\cite{Cao:2017drk,SNO:2021xpa}.
Selecting from among the wide range of available solvents for the LS, we used linear alkylbenzene~(LAB, $\rm C_6H_5C_nH_{2n+1}, where~ n=10-16$), an increasingly popular choice
because of its non-toxicity and cost-effectiveness.
The LAB-LS solution included 3~g/L of 2,5-diphenyloxazole ($\rm C_{15}H_{11}NO$) dissolved in the solvent for fluorescence and 30~mg/l of $p$-bis($o$-methylstyryl) benzene ($\rm C_{24}H_{22}$) as a wavelength shifter~\cite{Adhikari:2017esn}.

The NaI--PS--LS detector was coupled to a three-inch PMT made by Hamamatsu (R6233) on each end.
This PMT has a gain of $2.7\times10^5$, a spectral response range of 300 to 650~nm, and a peak wavelength of 420~nm\footnote{https://www.hamamatsu.com/eu/en/product/optical-sensors/pmt/pmt\_tube-alone/head-on-type/R6233.html}.
The NaI--PS--LS detector and the two PMTs were coupled 
using a silicone optical grease\footnote{We used Eljen Technology EJ-550, with an index of refraction of 1.46.}.
The complete assembly was then installed in a dark box with lead shielding averaging 50~mm in thickness.

We collected gamma calibration data regularly using $^{241}$Am (59.5~keV) and $^{60}$Co (1173.2 and 1332.5~keV) sources
as well as several sets of environmental radioactivity data between these calibration runs.
The signals from both PMTs were digitized using a 500-megasample-per-second fast analog-to-digital converter~(fast ADC) with a 2.5-V dynamic range.
Whenever the signal exceeded 20 ADC counts (greater than 12~mV) in both PMTs within a 200-ns time window, it triggered the recording of 8 microseconds of the PMT waveforms.
Raw data were collected in binary format and later converted into an analysis format\footnote{https://root.cern/}
for further analysis by a custom post-processing algorithm.

\section{Methods of analysis}

Because scintillation properties differ according to particle type and target material~\cite{psd},
a pulse shape analysis can be used to identify the various radioactive particles from the raw waveforms.
The main goal of the analysis is to develop an observable that characterizes
features in these raw waveforms.
Typically, observables are constructed by taking advantage of the pulse's decay time and the energy deposited.
In practice, the raw data are converted into separate data sets that contain event-wise observables such as average time and total charge.

Here, we define an energy-dependent observable denoted $\rm Q_{\mathrm{tot}}$ as the area integrated under the raw pulse
and a discrimination observable denoted $\rm T_{\mathrm{ave}}$ as the amplitude-weighted average of the pulse times.
These two observables are defined as
\begin{equation}
 \rm Q_{\mathrm{tot}}~(\mathrm{ADC})   = \sum_{\rm i}^{\rm bins}{q_i}+\sum_{\rm j}^{\rm bins}{q_j},~~~~~~~~~T_{\mathrm{ave}}~(\mathrm{ns})   = \frac{1}{2} \bigg( \frac{\sum_{i}^{bins}{q_it_i}}{\sum_{i}^{bins}{q_i}} + \frac{\sum_{j}^{bins}{q_jt_j}}{\sum_{j}^{bins}{q_j}}\bigg),
\end{equation}
where $\rm q_i (q_j)$ and $\rm t_i(t_j)$ are the $\rm i^{th}(j^{th})$ bin amplitude and its time
for PMT-1(PMT-2), respectively,
and the number of bins goes from the trigger time at approximately 2500~ns to the next 500~ns with an increment of 2~ns.

The $\rm Q_{\mathrm{tot}}$ parameter is converted to an energy value after calibration with several known full peaks of gamma rays.
The $\rm T_{\mathrm{ave}}$ variable, representing the mean time of a pulse, is strongly correlated with the pulse's decay time,
but the value it measures is slightly shorter than the decay time.
Because it only takes into account the shape,
the variable is only weakly dependent on energy.
Therefore, it introduces little energy bias due to event selection.

\begin{figure}[!htb]
  \begin{center}
      \includegraphics[width=0.90\textwidth]{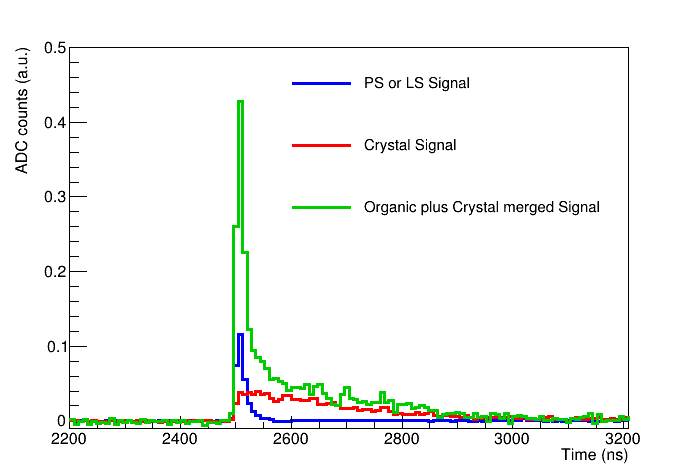}
  \end{center}
  \caption{Typical signal waveforms. The organic scintillator waveform (blue) shows a narrow pulse occurring within a short time, whereas that of the crystal (red) shows a relatively wide pulse. The waveform that includes both the organic scintillation signal and the crystal signal is shown in green. The range for the calculation of $\rm T_{\mathrm{ave}}$ and total charge is from $\sim$2500 to $\sim$3000~ns.}
  \label{wave}
\end{figure}

There are three signal waveforms in this detector configuration.
A pulse from the PS or LS signal is narrow and tall, whereas that from the crystal is wide and short for the same energy deposited.
A combined pulse contains proportionate amounts of narrow and wide components.
Examples of pulses are presented in Fig.~\ref{wave}.

\begin{figure}[!htb]
  \begin{center}
      \includegraphics[width=0.90\textwidth]{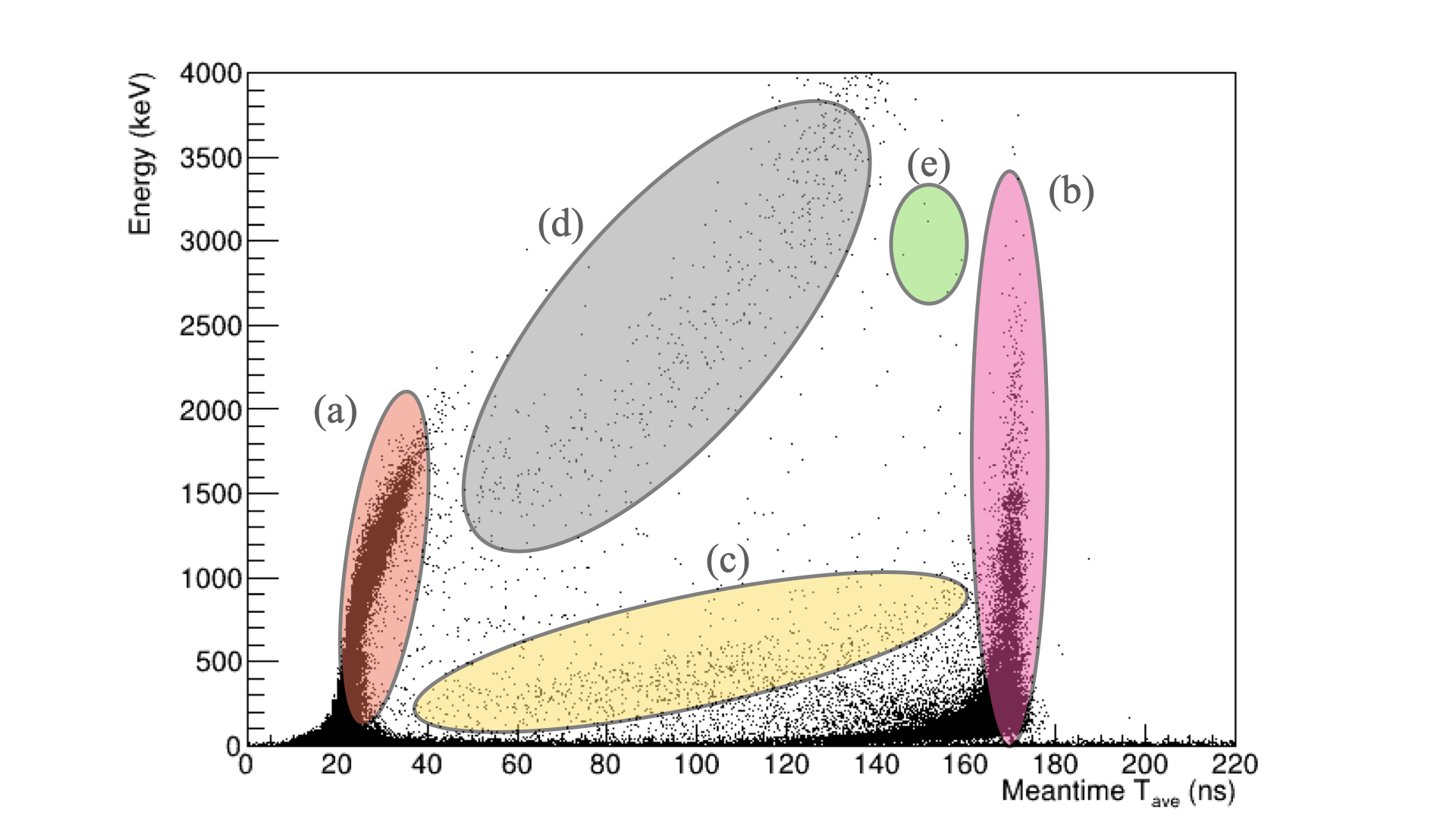}
  \end{center}
  \caption{Pulse shape parameter space. Energy is plotted as a function of $\rm T_{\mathrm{ave}}$ for events.
    Organic scintillator signals are in region (a), and crystal signals are in (b).
    Signals for crystal plus organic scintillator combined are in regions (c) and (d).
    Crystal alpha events appear in region (e).
  }
  \label{leg}
\end{figure}

Figure~\ref{leg} shows a typical pulse shape parameter space, energy as a function of the variable $\rm T_{\mathrm{ave}}$.
The y-axis has already been calibrated from $\rm Q_{\mathrm{tot}}$ (ADC) to energy (keV), and
the x-axis has been adjusted so that the value of the observable starts at 0~ns.
Region (a) contains the organic scintillation light, and (b) the crystal scintillation light.
In regions (c) and (d), we find the combined signals of LS or PS plus crystal.
Region (d) represents the particular case when an energetic cosmic-ray muon passes through the detector.
Finally, we expect to find the alpha decay population in region (e).
An alpha decay within a crystal can generate a signal waveform with a slightly sharper leading edge and a steeply dropping trailing edge.
Therefore, it will be located in a $\rm T_{\mathrm{ave}}$ region that is approximately $20$--$30$~ns shorter than the crystal beta/gamma region.
After the particles have been identified, we apply an event selection process that enables a spectral analysis and
the rate estimation.

\section{Experiment and results}

\begin{figure*}[!htb]
  \begin{center}
      \includegraphics[width=0.32\textwidth]{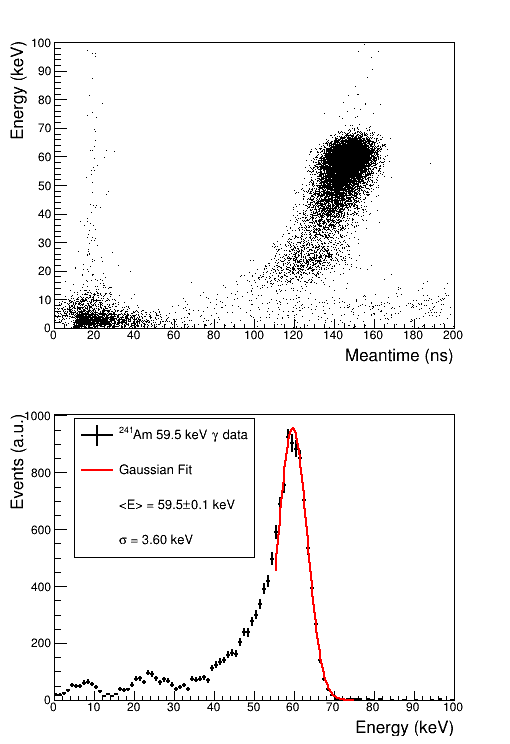}
      \includegraphics[width=0.32\textwidth]{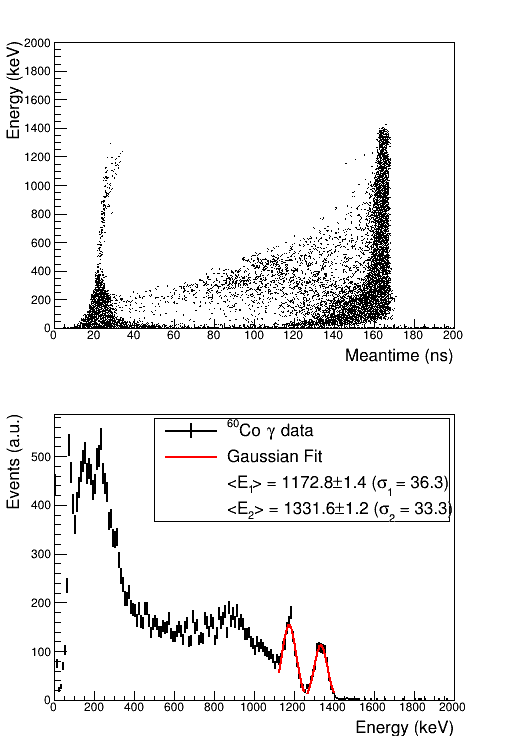}
      \includegraphics[width=0.32\textwidth]{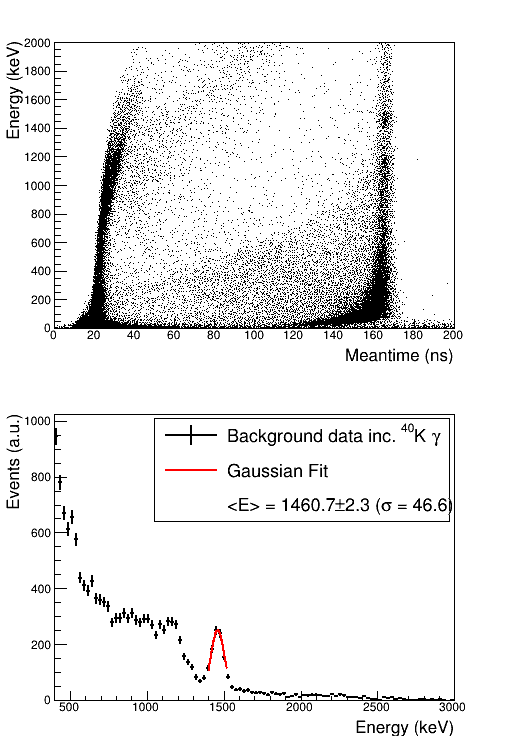}
  \end{center}
  \caption{Calibration of the crystal detector. Using external gamma-ray sources, the detector was calibrated for a wide range of crystal energies.
    The left column shows low-energy calibration, using an $^{241}$Am source; the middle and right columns show higher-energy calibrations, using $^{60}$Co and $^{40}$K, respectively.
    After selecting the crystal region using the $\rm T_{\mathrm{ave}}$ value, the full peaks identified were fitted with a Gaussian function, from which
    the mean energy $\rm \langle E \rangle$ and peak width $\sigma$ were extracted. Plots shown here are with the energy calibration applied. }
  \label{cal}
\end{figure*}

Calibrations of the detector were performed with known external gamma energy deposited in the crystal scintillator, including $^{241}$Am for 59.5~keV,  $^{60}$Co for 1173.2~keV and 1333.5~keV, and $^{40}$K for 1460.8~keV.
We did not calibrate the organic scintillators separately as it is not easy to record the full peaks for the gamma-ray sources.
Therefore, the energy of each event was referenced to the crystal spectrum regardless of its mean pulse time value.
In each calibration and background-only run campaign, we measured the positions and widths of the full peaks in ADC units
in order to use them as crystal quality parameters as a function of time.
The calibration results are shown in Fig.~\ref{cal}.
The pulse shape discrimination analysis provides a clear separation of crystal events from organic scintillator events.
The crystal linearity is well reproduced in the region between 100~keV and 10,000~keV, but
nonlinear behavior emerges below 100~keV, primarily because of the low number of photons recorded.

\begin{figure*}[!htb]
  \begin{center}
      \includegraphics[width=1\textwidth]{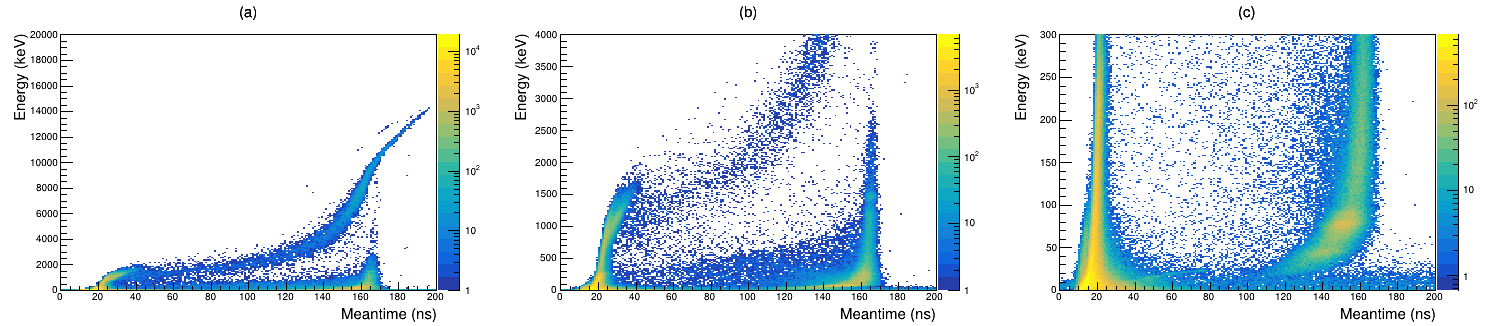}
  \end{center}
  \caption{
    Energy distribution as a function of $\rm T_{\mathrm{ave}}$ for the environmental background data. The left plot (a) shows the data for energies from 0~keV up to 20000~keV (high energy), the middle plot (b) from 0~keV up to 4000~keV (midium energy), and the right plot (c) from 0~keV up to 300~keV (low energy). High mean-time values (around 170~ns) are for crystal-originated signals, whereas lower values (around  20~ns) are for organic scintillator signals. The values in between are for the signals of crystal and organic scintillators combined. The energy is calibrated based on the crystal signals. The organic scintillator signals due to high-energy cosmic-ray muons become saturated earlier than that of the crystal, and therefore at the higher energies, the data curve toward higher $\rm T_{\mathrm{ave}}$ values.  The slanting trend of the combined signals is due to the difference in light yield between the organic scintillators and the crystal.
  }
  \label{meantimepsenergy}
\end{figure*}

Figure~\ref{meantimepsenergy} shows the classification of various interactions for a wide range of energy depositions in background-only data.
In all energy regions, the charge-weighted-mean-time parameter shows a clear separation between organic scintillator and crystal scintillator signals; the plastic and liquid signals are located in the region around 20~ns, whereas crystal signals are in the region around 170~ns.
The mismatch between these numbers and the actual decay times is due to limitations of the data acquisition system (jitter and shaping)
and the specific area integration window selected.
The data for organic scintillator events begin to curve toward higher values of $\rm T_{\mathrm{ave}}$ earlier than do those for the crystal events
because the waveform can become saturated earlier with peaks that are relatively sharp and tall, causing $\rm T_{\mathrm{ave}}$ to shift to a higher value.

\begin{figure*}[!htb]
  \begin{center}
    \begin{tabular}{cc}
      \includegraphics[width=0.5\textwidth]{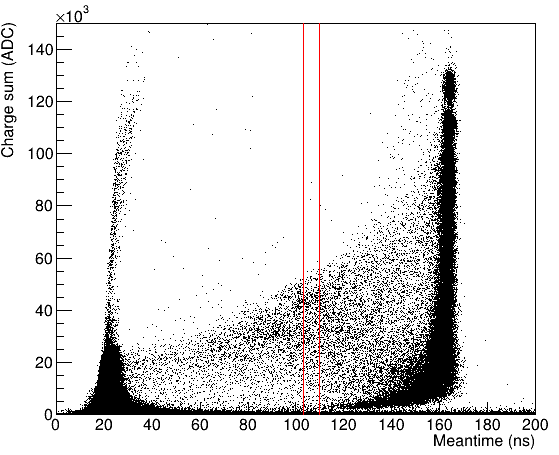} &   \includegraphics[width=0.5\textwidth]{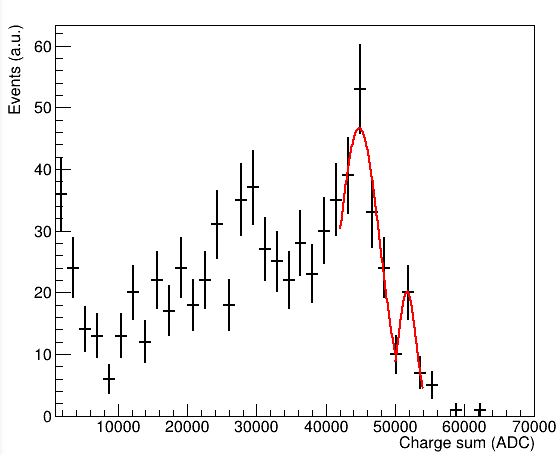} \\
      (a) &   (b) 
    \end{tabular}
  \end{center}
  \caption{Charge spectrum for $^{60}$Co full peaks, with energies deposited partly in the PS and partly in the crystal.
    Selection of $103~\mathrm{ns}<T_{\mathrm{ave}}<110~\mathrm{ns}$ (vertical red lines of plot (a)) was applied to the source data.
    Two distinct full peaks (plot (b)) are identified in this spectrum.
  }
  \label{partial}
\end{figure*}
Between two distinct signals, e.g., in the $103~\mathrm{ns}<T_{\mathrm{ave}}<110~\mathrm{ns}$ region, there are populations of events that produce both waveforms, such as
when external gamma-ray photons undergo Compton scattering at the PS or LS and then deposit their remaining energy
in the crystal. Full peaks for this class of events can also be identified, as shown in Fig.~\ref{partial}.
Signals from events of this class can be further identified as PS+crystal or LS+crystal combinations
as the PS and LS differ in their light output, creating the combined signals appearing in different positions in the $\rm T_{\mathrm{ave}}$ space.
Additionally, cosmic-ray muons are easily distinguishable in the higher-energy region, for which combined signals are recorded in most cases.
However, their signal amplitudes are too great under the current dynamic range settings, and therefore they have high saturation impact.
Alpha decay events are not clearly identifiable in this detector configuration because of the small size of the environmental background data set and the low level of $^{210}$Po contamination within the crystal.

\begin{figure*}[!htb]
  \begin{center}
      \includegraphics[width=0.95\textwidth]{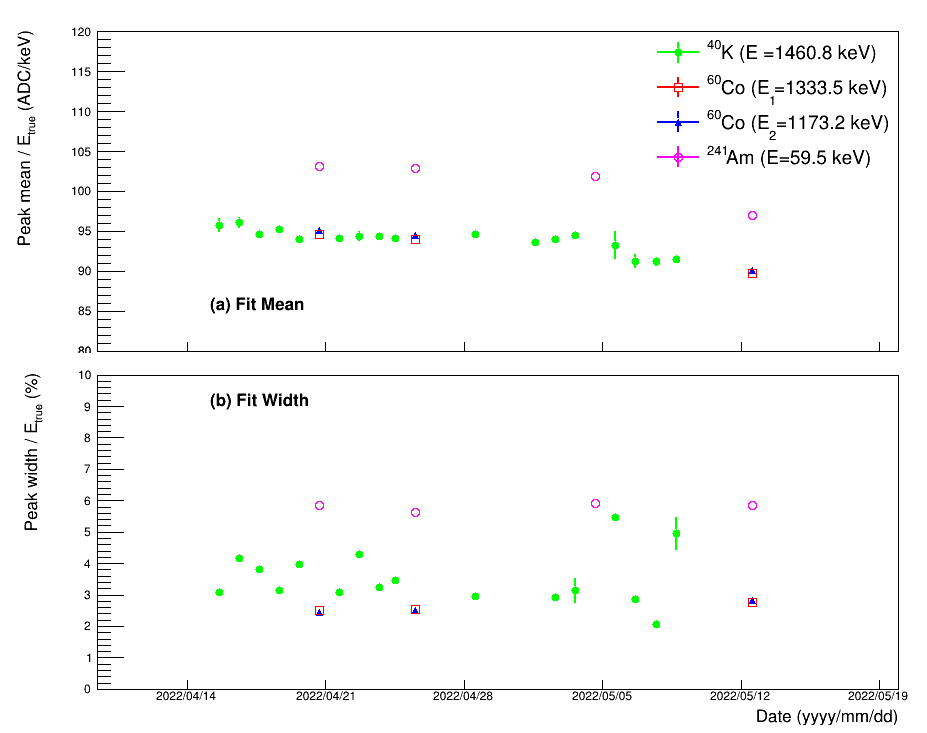}
  \end{center}
  \caption{Mean positions and widths of gamma-ray peaks as a function of time.
    For the four gamma-ray lines, the Gaussian-fit mean positions (a) and widths (b)
    normalized for their respective true energies were monitored for one month.
    The upper plot assesses changes in the measured gain over this period, and the lower plot assesses changes in the measured
    peak resolutions. Both plots show that the values remain stable. The error bars are statistical only.}
  \label{timeseries}
\end{figure*}

We investigated the stability of the detector by monitoring the positions and widths of the full peaks for gamma rays.
The ratios of the peak positions to the true gamma energies were monitored to assess the gain stability,
and the percent resolution (defined as the fitted width divided by the mean peak position) was monitored to assess light loss.
Both parameters were measured approximately once per week for a one-month period.
We found no notable change in these two indicators (shown in Fig.~\ref{timeseries}), and
the individual components are within 3\% of the constant line fitted to the measurement points.
We therefore conclude that the phoswich configuration protects the crystal from contact with air.

\section{Conclusions}

In summary, we have constructed a phoswich scintillation detector consisting of one core crystal and two outer layers of organic scintillators. We investigated its radiation detection capability using pulse shape discrimination analysis. Various types of radioactive interactions were identified in the results from this detector configuration. A clear separation of crystal signals from PS/LS signals was achieved through the use of an amplitude-weighted-mean-time parameter. Additionally, the combined pulse that includes both organic and inorganic signals was extensively studied.

The NaI(Tl) crystal's light output weakens when it comes into contact with water molecules owing to the change in transmittance at the surface. Thus, the crystal needs to be tightly encapsulated to avoid exposure to air, and therefore we encase the crystal in PS and then additionally in LS. With these outer organic scintillator layers, the crystal is doubly protected. The light from the crystal can travel through the layers to reach PMTs, and the layers themselves act as independent sub-detectors. This makes possible a radiation veto for external radiation that reaches the bulk of the crystal. Furthermore, a more accurate determination of the position of an initial interaction involving an external gamma ray can be made when Compton scattering occurs in the organic scintillator and then the scattered photon deposits its full energy in the crystal.

The detector assembly was tested for a month using several radiation sources, during which the light output of the crystal was monitored indirectly by measuring the positions and widths of the full gamma-ray peaks. We found that the assembly showed no substantial variation in either of these indicators at the low energy or the high energies. The Gaussian-fit positions and widths of the gamma peaks were within 3\% of a constant line for the test period.

A more complete understanding of this triple phoswich detector concept will require further study. As particular examples, a method could be developed  to minimize the loss of crystal-produced photons during their travel to the PMTs, and a more sophisticated coupling between crystal and PS needs to be implemented. Ongoing developments include a similar phoswich detector in which the crystal surface is artificially contaminated with $^{210}$Po and the crystal's alpha decays are identified in situ. This would provide a better understanding of the radioactive background of the crystal, especially in rare decay experiments. Overall, this improvement to NaI(Tl) crystal encapsulation contributes to the advance of gamma-ray spectroscopy and related medical applications. 

\begin{acknowledgements}
  This research was supported by the Chung-Ang University Research Scholarship Grants in 2021 and
  by a National Research Foundation of Korea~(NRF) grant funded by the Korean government~(MSIT) (No. 2021R1A2C1013761).
\end{acknowledgements}



\end{document}